\documentclass[twocolumn,prb,aps,showpacs]{revtex4}
\usepackage{graphics}
\usepackage{dcolumn}
\usepackage{bm}
\usepackage{amsmath}
\usepackage{amssymb}
\usepackage{epsfig}
\usepackage{pictex}
\begin{document}
\title{ Orbital Selective Superconductivity in Two-Orbital Asymmetric t-J Models }
\author{Feng Lu$^{1,2}$,  Cui-Zhi Wang$^{1,2}$,  Wei-Hua Wang$^{3}$,
and Liang-Jian Zou$^{1
\footnote{Correspondence author, Electronic mail: zou@theory.issp.ac.cn}}$}
\affiliation{\it
 $^1$ Key Laboratory of Materials Physics, Institute of Solid State Physics,
      Chinese Academy of Sciences, P. O. Box 1129, Hefei 230031, China  \\
\it $^2$  Graduate School of the Chinese Academy of Sciences,
          Beijing 100049, China \\
\it $^3$ Department of Electronics, College of Information Technical
         Science, Nankai University, Tianjin 300071, China}
\date{2008-03-06}

\begin{abstract}
We present the zero-temperature superconducting (SC) ground states
of the two-orbital asymmetric $t-J$ model on a square lattice by
means of the auxiliary-boson approach. Besides the two-gap SC phase,
we find an orbital selective SC (OSSC) phase, which is
simultaneously SC in one orbit and normal in another orbit. The
novel OSSC phase is stable only for sufficient asymmetric degree in
orbital space and doping concentration.
The pairing symmetry of the SC phase is $s$-wave-like in most doping
regime, against the $d$-wave symmetry of the single-orbital $t-J$
model in a square lattice.
The implication of the present scenario on the iron-based and other
multi-orbital superconductors is also discussed.

\end{abstract}


\maketitle

\section{Introduction}
Searching high-T$_{c}$ superconducting states in correlated electron
systems has been one of the central topics in condensed matter
physics for several decades \cite{Lee0}. With the development of
synthesis technique, more and more new superconducting (SC)
compounds have been discovered, such as UGe$_{2}$
\cite{Saxena,Huxley} and URhGe \cite{Aoki} with the coexistence of
ferromagnetism and SC, and LaO$_{0.9}$F$_{0.1}$FeAs with the
anomalous Hall coefficient and magnetoresistance in the SC
\cite{Hosono, Yang}, etc.
Among the fantastic properties of these compounds, the multi-gap SC
nature is one of the most interested. The multi-gap SC characters
have been found in a number of compounds, for example, in nodal
structure Sr$_2$RuO$_4$\cite{Maeno,Mackenzie}, in $s$-wave SC
NbSe$_2$\cite{Yokoya, Boaknin} and MgB$_2$
\cite{Canfield,Blumberg,Iavarone}, in heavy-fermion SC
Ce$_{1-x}$La$_x$CoIn$_5$ \cite{Tanatar,Flouquet}, and in d-wave SC
PbBi$_{2}$Sr$_{2}$CuO$_{6+x}$\cite{Boyer} and Bi2212 \cite{Lee}.
Recently, a generic two-gap hypothesis is also proposed for the
cuprate SC \cite{Sawatzky}. Further, in the iron-based SC discovered
recently \cite{Hosono}, some experiments also showed that
LaO$_{1-x}$F$_{x}$FeAs is two-gap SC \cite{Mandrus, Zhenggq}.
On the other hand, in the heavy-fermion SC Ce$_{1-x}$La$_x$CoIn$_5$
\cite{Tanatar,Flouquet}, it is found recently that there coexist
uncondensed carriers and the SC carriers, and a fraction of
electrons on the Fermi surfaces does not participate in SC,
displaying unusual SC characteristics in Ce$_{1-x}$La$_x$CoIn$_5$.
Since these compounds share the same properties in the electronic
structures, {\it i.e.}, the multi-orbital or multi-band character
\cite{Agterberg,Liu2,Barzykin}, it naturally arises the question
that whether the condensed and the uncondensed phases generally
coexist in the multi-orbital SC compounds ?
%
%
%

In the past decade, it has been found that numerous correlated
electron compounds are multi-orbital, and many unique features are
contributed from the orbital degree of freedom, such as colossal
magnetoresistance and complex orbital ordered phases in doped
manganites\cite{Imada}, the exotic magnetism in the $f$-electron
systems \cite{Santini}, and the debated orbital selective Mott
transition (OSMT) in Sr$_{2-x}$Ca$_{x}$RuO$_{4}$
\cite{Anisimov,Nakatsuji,Lee2,Wang,Dai}, $etc.$
%
%
The orbital degree of freedom on the SC plays an important role in
the SC pairing symmetry of the multi-orbital systems
\cite{Takimoto,Takimoto2,Mochizuki,Kubo}.
On the other hand, Liu $et$ $al.$\cite{Liu} proposed that there may
exist an interior gap superfluidity in a two-orbital system, in
which the pairing interaction carves out a gap within the interior
of a large Fermi ball, while the exterior surface remains gapless;
and it is a simultaneously SC and normal state at zero temperature.
Nevertheless, the role of orbital degree of freedom on the pairing
symmetry and SC condensation is far from well understood.
%
%
So, much effort is still needed to uncover the SC nature in strongly
correlated multi-orbital compounds.

With this motivation, and stimulated by searching for a new state of
matter, we study the SC properties of the strongly correlated
multi-orbital systems in this paper. We extend the single-orbital
$t-J$ model to the two-orbital $t-J$ model, and utilize the
auxiliary-boson method \cite{Coleman} to obtain the phase diagram of
the system on a square lattice.
We find that for given strong interactions between electrons, the
electrons may bind as many paired quasiparticles as possible in one
orbit due to the symmetry breaking in the orbital space. These
quasiparticles condense into a coherent state at low temperature,
and the residual unpaired electrons in the other orbit constitute a
separated normal fluid, forming the orbital selective SC (OSSC)
phase. The OSSC phase is the stablest, against to the normal phase
and the full-gapped SC ones, in proper parametric and doping regime.
We also obtain the critical points of the formation of OSSC phase
for various parameters of the two-orbital $t-J$ model, which may
shed light on finding this phase in the strongly correlated
compounds. The possible application of the present theory on the
Fe-based SC LaO$_{1-x}$F$_{x}$FeAs is also discussed.

The rest of this paper is organized as follows. In Sec.II, we
describe the two-orbital $t-J$ model and the framework of the
auxiliary-boson mean field approach. And then we present the
detailed formulation of the two-orbital SC. In Sec.III, we present
the phase diagram of the SC in the two-orbital system. The
conditions for the formation of the OSSC phase and its possible
application on Ce$_{1-x}$La$_x$CoIn$_5$ and Fe-based SC are
discussed. The last part is devoted to the summary.

\section{HAMILTONIAN AND METHOD}
\label{secmodel}

%
The Hubbard model effectively describes the electronic correlations
in the strongly interacting systems. In the single-orbital situation
and in the strong correlation limit, the Hubbard model is reduced to
an effective model describing the low-energy processes, {\it i.e.}
the single-orbital $t-J$ model\cite{Anderson,Baskaran}.
In the strongly correlated multi-orbital system\cite{Castellani}, we
naturally extend the single-orbital $t-J$ model to the multi-orbital
t-J model described by the Hamiltonian
\begin{eqnarray}
  H&=&P[H_{t}+H_{J}]P.
  \nonumber\\
\end{eqnarray}
with a kinetic energy part $H_{t}$
\begin{eqnarray}
  H_{t}&=&\sum_{<i j> m m^{\prime}\sigma}t_{i j}^{m m^{\prime}}
          ( c^{\dagger}_{i m \sigma} c_{j m \sigma}+h.c )
          +\sum_{i \sigma}E_{m} n_{i m \sigma}
  \nonumber\\
\end{eqnarray}
and a general superexchange coupling $H_{J}$ \cite{Castellani}.
\begin{eqnarray}
  H_{J}&&=-\sum_{<i j>\sigma}\sum_{n n^\prime m}
          c^{\dagger}_{i n \sigma} c_{i n^\prime \sigma} [
          J^{n m n^\prime m}_{1} (n_{j m \bar{\sigma}}
          + c^{\dagger}_{j m \bar{\sigma}} c_{j m\sigma})
  \nonumber\\
 &&+(J^{n m n^\prime m}_{2}n_{j \bar{m} \bar{\sigma}}
   +J^{n m n^\prime \bar{m}}_{2}
     c^{\dagger}_{j \bar{m} \bar{\sigma}} c_{j m \sigma})
  \nonumber\\
 && + (J^{n m n^\prime m}_{3} n_{j \bar{m} \sigma}
    +  J^{n m n^\prime \bar{m}}_{3} c^{\dagger}_{j \bar{m}
    \sigma} c_{j m \sigma})
  \nonumber\\
  &&- (J^{n m n^\prime m}_{4} c^{\dagger}_{j \bar{m} \bar{\sigma}}
       c_{j \bar{m} \sigma}
     - J^{n m n^\prime \bar{m}}_{4} c^{\dagger}_{j \bar{m}
     \bar{\sigma}} c_{j m \bar{\sigma}})
  \nonumber\\
  && +( J^{n m n^\prime \bar{m}}_{5}
        c^{\dagger}_{j m \bar{\sigma}} c_{j \bar{m} \bar{\sigma}}
      - J^{n m n^\prime \bar{m}}_{5}
        c^{\dagger}_{j m \bar{\sigma}} c_{j \bar{m} \sigma})],
  \nonumber\\
\end{eqnarray}
where $c^\dag_{i m \sigma}(c_{i m \sigma})$ is the creation
(annihilation) operator of the electron at site $i$ with orbit-m and
spin $\sigma (=\uparrow, \downarrow)$; and $\bar{m}$
($\bar{\sigma}$) denotes the orbit (spin) other than m ($\sigma$).
$n_{i m \sigma}(=c^\dag_{i m \sigma}c_{i m \sigma})$ is the electron
number operator. E$_{m}$ denotes the crystalline field level of the
orbit-$m$. The operator $P$ projects out the states of multiple
occupancy at each site.
Therefore, every site is either spin 1/2 or a vacancy. $t_{i j}^{n
m}$ denotes the hopping integral from the $m$ orbit at site j to the
$n$ orbit at site i, and only the nearest-neighbor hopping is taken
into account. In this paper, we define $t_{i j}^{1 1}=t=1$ as the
energy unit. The parameter J$^{n m n^\prime m^\prime}$ is associated
with the hopping integrals t$^{n m}$, the on-site and inter-orbital
Coulomb interactions $U$ and $U^{\prime}$, and the Hund's coupling
via,
\( J^{n m n^\prime m^\prime}_{1}={t_{i j}^{n m} t_{i j}^{n^\prime
m^\prime}U}/{(U^{2}-J^{2})} \),
\( J^{n m n^\prime m^\prime}_{2}={t_{i j}^{n m} t_{i j}^{n^\prime
m^\prime}U^\prime}/{(U{^\prime}^{2}-J^{2})} \),
\( J^{n m n^\prime m^\prime}_{3}={t_{i j}^{n m} t_{i j}^{n^\prime
m^\prime}}/{(U{^\prime}-J)} \),
\( J^{n m n^\prime m^\prime}_{4}={t_{i j}^{n m} t_{i j}^{n^\prime
m^\prime} J}/{(U{^\prime}^{2}-J^{2})} \),
\( J^{n m n^\prime m^\prime}_{5}={t_{i j}^{n m} t_{i j}^{n^\prime
m^\prime} J}/{(U^{2}-J^{2})} \).
In what follows, considering the spin rotational symmetry, we adopt
the relationship $U=U'+2J$\cite{Castellani}, and the system is on a
square lattice.

Eq.(1) is a general multi-orbital $t-J$ Hamiltonian with the
electron/hole filling not far from 1/4. In the absence of the Hund's
coupling and let the hopping integrals be isotropic for all of the
orbits, Eq.(1) reduces to a supersymmetric SU(4) $t-J$ model
\cite{Fujii, Schlottmann}.
If the crystalline field splitting, $E_{\Delta}$=$E_{2}-E_{1}$, is
sufficiently large, only the orbit/band m=1 is occupied, and Eq.(1)
reduces to the traditional single-orbital $t-J$ model.
For clarify, we concentrate on the two-orbital situation (m=1,2)
without only the off-diagonal hoppings (t$^{12}$=0) and the Hund's
coupling ($J=0$). Thus the effective Hamiltonian $H_{eff}$ in Eq.(1)
becomes,
\begin{eqnarray}
  H_{t}&=&\sum_{<i j> m \sigma}t_{i j}^{m m} ( c^{\dagger}_{i m \sigma}
          c_{j m \sigma}+h.c )
          +E_{\Delta} \sum_{i \sigma}(n_{i 1 \sigma}-n_{i 2 \sigma})
\nonumber\\
  H_{J}&=&-\sum_{<i j> \sigma} [J_{1} n_{i 1 \sigma} n_{j 1 \bar{\sigma}}
          +J_{2} n_{i 2 \sigma} n_{j 2 \bar{\sigma}}+(J_{1}+J_{2})
\nonumber\\
   && (n_{i 1 \sigma}
      n_{j 2 \bar{\sigma}} +n_{i 1 \sigma} n_{j 2 \sigma})-
      J_{1} c^{\dagger}_{i 1 \sigma} c_{i 1 \bar{\sigma}}
      c^{\dagger}_{j 1 \bar{\sigma}} c_{j 1 \sigma}
\nonumber \\
   && - J_{2} c^{\dagger}_{i 2 \sigma} c_{i 2 \bar{\sigma}}
      c^{\dagger}_{j 2 \bar{\sigma}} c_{j 2 \sigma}- 2 J_{3}
      (c^{\dagger}_{i 1 \sigma} c_{i 2 \sigma} c^{\dagger}_{j 2 \sigma}
      c_{j 1 \sigma}
\nonumber \\
   && - c^{\dagger}_{i 1 \sigma} c_{i 2 \bar{\sigma}}
      c^{\dagger}_{j 2 \bar{\sigma}} c_{j 1 \sigma}) ],
\end{eqnarray}
with $J_{1}=(t^{11})^{2}/U$, $J_{2}=(t^{22})^{2}/U$ and
$J_{3}=t^{11} t^{22}/U$.
%
%
To enforce the single occupation constraint at each site, we employ
the auxiliary-boson mean-field approximation \cite{Kotliar} on
Eq.(4).

Within the slave-boson representation, the Hamiltonian (4) is
rewritten in terms of the projected fermion operators
$Pc^{\dagger}_{im\sigma}P$ and $Pc_{im\sigma}P$, which rule out the
double and multiple fermion occupancies at every site, as well as
the slave boson operators.
The constrained Hilbert space ($S$) of each site $i$ is
\begin{eqnarray}
  S_{i}=\{ {\mid 1, \uparrow >}, {\mid 1, \downarrow >},
   {\mid 2, \uparrow >},{\mid 2, \downarrow >}, {\mid 0,0 >} \},
\end{eqnarray}
including the states of spin-up, spin-down in the orbit-1, and those
in the orbit-2, together with vacancy state, respectively.
  The present constrained spin-orbital formulation resembles to the
4-fold degenerate state of pseudo-angular momentum $j$=$3/2$
proposed by Coleman\cite{Coleman}, if we define $\mid f^{0}
>$=$\mid 0,0 >$, ${\mid f^{1}: 3/2, -3/2
>}$=${\mid 1, \uparrow >}$, ${\mid f^{1}: 3/2, -1/2 >}$=${\mid 1,
\downarrow >}$, ${\mid f^{1}: 3/2, 1/2 >}$=${\mid 2, \uparrow >}$,
and ${\mid f^{1}: 3/2, 3/2 >}$=${\mid 2, \downarrow >}$. One obtains
\( P c^{\dagger}_{i m \sigma} P = f^{\dagger}_{i m \sigma}. \)
The boson operator $b_{i}^{\dagger}$ creates an empty occupation
state at the $i$th site, and the fermion operator $f_{i m
\sigma}^{\dagger}(f_{i m \sigma})$ creates (annihilates) a slaved
electron at site $i$ with orbit-m and spin $\sigma (=\uparrow,
\downarrow)$.

In the present situation, we define the SC order parameters as
\begin{eqnarray}
  \Delta^{\sigma, m}_{x/y}&=&
  <f^{\dagger}_{i m \sigma} f^{\dagger}_{j m \bar{\sigma}} - f^{\dagger}_{
  i m \bar{\sigma}} f^{\dagger}_{j m \sigma}>,
        (j=i \pm \hat{x}/\hat{y})
    \nonumber\\
  {P_{m \sigma}}&=&<f^{\dagger}_{i m \sigma} f_{j m \sigma}>,
    ~
  {\bar{n}_{i m \sigma}}= <f^{\dagger}_{i m \sigma} f_{i m
  \sigma}>
\end{eqnarray}
Considering the asymmetry of the orbital space and the Fermi surface
topology, the $p$-wave SC order parameters vanish.
In the mean-field or saddle-point approximation, we have
$<b^{\dagger}_{i}b_{j}>=\mid b\mid ^{2}=\delta$, and the electron
filling n=1-$\delta$.
Thus, we obtain the slave-boson mean-field Hamiltonian
\begin{eqnarray}
     H&&= E_{0}+\sum_{{\bf k} \sigma} \epsilon_{{\bf k}1\sigma}
         f^{\dagger}_{{\bf k}1
         \sigma}f_{{\bf k}1\sigma} -\sum_{{\bf k} \sigma} (\Delta^{\sigma 1
         }_{\bf k}f^{\dagger}_{{\bf k}1\sigma}
         f^{\dagger}_{{\bf -k}1\bar{\sigma}} + h.c)
\nonumber\\
      &&+\sum_{{\bf k} \sigma} \epsilon_{{\bf k}2\sigma}f^{\dagger}_{{\bf
      k}2\sigma}f_{{\bf k}2\sigma} -\sum_{{\bf k} \sigma}
      (\Delta^{\sigma 2}_{\bf k}f^{\dagger}_{{\bf k}2\sigma}
        f^{\dagger}_{{\bf -k}2\bar{\sigma}}+h.c)
\end{eqnarray}
with the constant energy
\begin{eqnarray}
  E_{0}&=& N \sum_{\sigma} [2 J_{1} \bar{n}_{1 \sigma} \bar{n}_{1 \bar{\sigma}}
          + \genfrac{}{}{}{}{J_{1}}{2} ( \mid \Delta^{\sigma 1}_{x} \mid ^{2}
          + \mid \Delta^{\sigma 1}_{y} \mid ^{2} )
  \nonumber\\
  &&  + 2 J_{1} P_{1 \sigma} P_{1 \bar{\sigma}}
      +2 J_{2} \bar{n}_{2 \sigma} \bar{n}_{2 \bar{\sigma}}
          + \genfrac{}{}{}{}{J_{2}}{2} ( \mid \Delta^{\sigma 2}_{x} \mid ^{2}
          + \mid \Delta^{\sigma 2}_{y} \mid ^{2} )
  \nonumber\\
  && + 2 J_{2} P_{2 \sigma} P_{2 \bar{\sigma}}
     + 2 ( J_{1} + J_{2} ) (\bar{n}_{1 \sigma}\bar{n}_{2 \sigma}+\bar{n}_{1
       \sigma}\bar{n}_{2 \bar{\sigma}})
\nonumber\\
   &&  + 4J_{3}(P_{1 \sigma}P_{2 \sigma}+P_{1 \sigma}P_{2 \bar{\sigma}})]
\end{eqnarray}
%
%
and the notations
\begin{eqnarray}
   \epsilon_{{\bf k}m\sigma}&=&-2[t_{m}\delta+(J_{m}P_{m\bar{\sigma}}
     +J_{3}P_{\bar{m}\sigma}+J_{3}P_{\bar{m}\bar{\sigma}})]\gamma_{\bf
     k}-\mu
\nonumber \\
     &&-2[2J_{m}\bar{n}_{m\bar{\sigma}}
     +(J_{1}+J_{2})(\bar{n}_{\bar{m}\sigma}+\bar{n}_{\bar{m}\bar{\sigma}})]+E_{\Delta},
  \nonumber\\
\Delta^{\sigma m}_{\bf k}&=&J_{m}
     [\Delta^{\sigma m}_{x}cos(k_{x}a)+\Delta^{\sigma m}_{y}cos(k_{y}a)],
  \nonumber\\
\gamma_{\bf k}&=&cos(k_{x}a)+cos(k_{y}a).
\end{eqnarray}

   Diagonalizing Eq.(8) through the Valatin transformation \cite{Valatin,Callaway}.
\begin{eqnarray}
  f_{{\bf k}m \uparrow}&=&U_{{\bf k} m}{\alpha}_{{\bf k} m}
                         +V_{{\bf k} m}\beta^{\dagger}_{{\bf k} m},
  \nonumber\\
  f_{{\bf -k}m \downarrow}&=&U_{{\bf k} m}{\beta}_{{\bf k} m}
                         -V_{{\bf k} m}\alpha^{\dagger}_{{\bf k} m},
\end{eqnarray}
with $U_{{\bf k} m}^{2}+V_{{\bf k} m}^{2}=1$, here $U_{{\bf k} m}$
and $V_{{\bf k} m}$ (m=1,2) are real and positive.
The slave-boson mean-field Hamiltonian becomes
\begin{eqnarray}
  H&=&\sum_{\bf k} [\xi_{{\bf k} 1}
        (\alpha^{\dagger}_{{\bf k} 1}\alpha_{{\bf k} 1} +
       \beta^{\dagger}_{{\bf k} 1}\beta_{{\bf k} 1} ) +
        \xi_{{\bf k} 2} (\alpha^{\dagger}_{{\bf k} 2}\alpha_{{\bf k} 2}
         +\beta^{\dagger}_{{\bf k} 2}\beta_{{\bf k} 2} ) ]
      \nonumber\\
      && + E_{0} +\sum_{\bf k}[(\epsilon_{{\bf k} 1}-
        \xi_{{\bf k} 1})+(\epsilon_{{\bf k} 2}-\xi_{{\bf k} 2})],
\end{eqnarray}
with the quasiparticle excitation spectrum,
\begin{eqnarray}
 \xi_{{\bf k} m \sigma}=\sqrt{\epsilon^{2}_{{\bf k}m\sigma}
    +4\mid \Delta^{\sigma m}_{\bf k}\mid ^{2}},
\end{eqnarray}
and the chemical potential $\mu$ is determined by $1-\delta=1/N
\sum_{i m \sigma}<f^{\dagger}_{i m \sigma}f_{i m \sigma}>$. The
order parameters P$_{m\sigma}$ and $\Delta^{\sigma m}$ satisfy the
self-consistent equations:
\begin{eqnarray}
  1&=&\genfrac{}{}{}{}{t^{2}_{m}}{UN}\sum_{\bf k}\genfrac{}{}{}{}{(cos(k_{x}a)
       +cos(k_{y}a))^{2}}{\xi_{{\bf k}m}}
  \nonumber\\
   P_{m\sigma}&=&\genfrac{}{}{}{}{1}{4N}\sum_{\bf k}(1-
    \genfrac{}{}{}{}{\epsilon_{{\bf k} m \sigma }}
  {\xi_{{\bf k} m \sigma }})\gamma_{\bf k}
\end{eqnarray}
for the $s$-wave SC. In the d-wave SC situation, these equations are
slightly different.

\section{RESULTS AND DISCUSSION}
\label{secmodel}

In this section, we present the essential properties of the SC and
the phase diagram of the two-orbital $t-J$ model.
Provided the two-orbital model is SU(2) symmetric in the spin space,
we would have the relations for the single occupation amplitude
$P_{m \uparrow}=P_{m \downarrow}=P_{m}$ and for the SC order
parameters, $\Delta^{\uparrow, m}_{x}=-\Delta^{\downarrow,
m}_{x}=\Delta_{m}$, and $\Delta^{\uparrow,
m}_{y}=-\Delta^{\downarrow, m}_{y}$ (m=1,2). This will greatly
simplify our discussion in what follows.

\subsection{Stable SC Phase and its Pairing Symmetry}
We firstly determine the stablest ground state of the two-orbital
$t-J$ model with n=1-$\delta$ in a square lattice through comparing
the groundstate energy of the possible GS candidates: the normal
state with $\Delta_{m}$=0, the $d$-wave symmetric SC phase with
$\Delta^{\uparrow, m}_{x}=-\Delta^{\uparrow, m}_{y}$, or the
$s$-wave-like symmetric SC phase with $\Delta^{\uparrow, m}_{x}
=\Delta^{\uparrow, m}_{y}$.
%
%
Throughout this paper, we adopt the relationship $J_{1}=0.1$,
$J_{2}=J_{1}R^{2}$ and $J_{1}=J_{3}R$, unless specific explanation.
Here the ratio of the hopping integrals is defined as R =
t$_{22}$/t$_{11}$. The larger R deviates from the unity, the larger
asymmetry degree of the orbital space is. By minimizing the GS
energy, we obtain phase diagrams of the systems for various
parameters, such as the doping concentration $\delta$, the exchange
parameters $J_{1}$ and the asymmetric ratio of the hopping
integrals, R, and the orbital level splitting, $E_{\Delta}$.
Our numerical results show that the energy of the $s$-wave-like SC
state is lower than that of the $d$-wave state for all the
situations we investigated.
%
%
Out of this reason, we focus on the s-wave like symmetric order
parameters of the multi-orbital $t-J$ model in the following.
%
%
The ground state strongly depends on the doping, the orbital
asymmetry ratio, the level splitting and the the superexchange
coupling strength, as shown the following phase diagrams. These
phase diagrams demonstrate that there exists a new SC phase that is
simultaneously SC and normal at zero temperature, analogous to the
interior gap superfluidity proposed by Liu $et$ $al.$ \cite{Liu}.

\subsection{Doping Dependence}

\begin{figure}[tp]
\vglue -0.6cm \scalebox{1.150}[1.15]
{\epsfig{figure=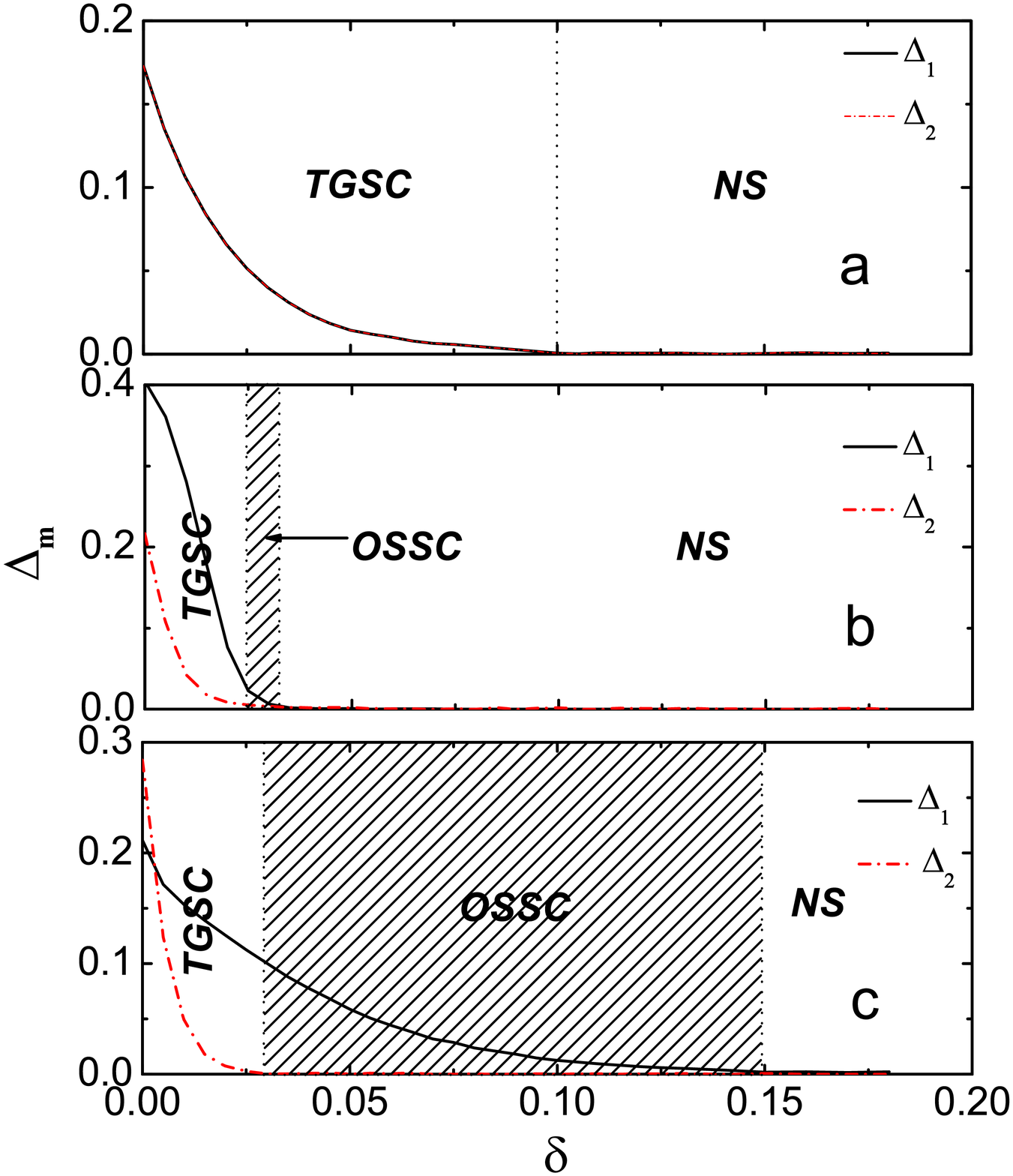,width=8cm,angle=0}} \caption{(Color
online) The order parameters $\Delta_{m}$ (m=1,2) as functions of
doping concentration $\delta$ for different situations, (a).$R=1$
and $\Delta=0$, (b).$R=0.3$ and $\Delta=0$, and (c).$R=1$ and
$\Delta=0.1$, respectively. }
\label{fig:fig1}
\end{figure}

We first present the evolution of the SC order parameters with the
doping concentration. The order parameters $\Delta_{m}$ as the
functions of doping $\delta$ in different parametric region are
distinctly different, as shown in Fig. 1.
When the system is SU(4) symmetric with $R=1$ and $E_{\Delta}=0$, it
is clearly seen that the SC order parameters of the two orbits,
$\Delta_{1}$ and $\Delta_{2}$, are equal.
With the increase of doping $\delta$, the SC order parameters
$\Delta_{1}$ and $\Delta_{2}$ simultaneously decrease and become
zero at a critical value $\delta_{c}=0.09$, which is smaller than
the critical value $\delta_{c}=0.32$ of the single-orbital $t-J$
model for SC phase.
So, in the two-gap SC (TGSC) phase with SU(4) symmetry, the SC
energy gaps of the two orbits are identical.
%
%

When the nearest-neighbor hoppings of the two orbits become
asymmetric, such as, in the situation of $R=0.3$ and $E_{\Delta}=0$,
we find that the order parameter of the orbit-1, $\Delta_{1}$,
deviates from that of the orbit-2, $\Delta_{2}$.
With the increase of $\delta$, the SC order parameter $\Delta_{2}$
in the narrow band firstly vanishes at a critical value,
$\delta_{c}=0.025$.
On the contrast, the SC order parameter $\Delta_{1}$ gradually
decreases and becomes zero at $\delta_{c}=0.035$. Thus in the doping
region $\delta<0.025$, the system is in the TGSC phase that both
orbits/bands are SC.
While in the region $0.025<\delta<0.035$, a novel phase appears, in
which the wide orbit is SC, while the narrow orbit is in the normal
phase. This new phase is called the orbital selective
superconductive (OSSC) phase, as seen the shaded region in Fig. 1b.
Since the degree of the broken symmetry is not large, such a phase,
analogous to the interior gap superfluidity\cite{Liu}, occurs only
in a very narrow region in the phase diagram.
%
%

The symmetry breaking arising from the crystal field splitting
E$_{\Delta}$ also leads to the stable OSSC phase. As shown in
Fig.1c, the OSSC phase in the situation $R=1$ and $E_{\Delta}=0.1$
is more robust.
In this situation, with the increase of $\delta$, both the SC order
parameters $\Delta_{m\sigma}$ decrease asynchronously, as seen in
Fig. 1c. $\Delta_{2}$ vanishes at $\delta_{c}=0.035$, and
$\Delta_{1}$ vanishes at $\delta_{c}=0.14$. There also exist three
different phases, the TGSC phase, the OSSC phase and the normal
ones. Among these phases, the presence of crystalline field
splitting favors the formation of the OSSC phase.
We notice that in Fig.1, when the orbital rotational symmetry is
broken, the SC order parameters at $\delta \rightarrow$ 0 are
considerably larger than those with orbital rotational symmetry,
demonstrating that the SC pairing strength may be enhanced by the
orbital asymmetry, as we seen in next subsection.
%
%

\subsection{Orbital Asymmetry Dependence}

  In fact, whether there exists the OSSC phase reflects the
orbital symmetry of the systems.
For $R=1$ and $E_{\Delta}=0$, the system has the SU(2) symmetry in
the orbital space. In other words, there is no difference between
the two orbits, their order parameters simultaneously decrease and
vanish. Therefore, the TGSC and the normal phases are the stablest
in this situation.
Whilst, in the situations of $R=0.3$, $E_{\Delta}=0$ and $R=1$,
$E_{\Delta}=0.1$, the rotation symmetry in the orbital space is
broken because of the inequivalence of the two orbits for the case
with $R=0.3$ and $E_{\Delta}=0$ or the case with $R=0$ and
$E_{\Delta}=0.1$. The asymmetry of the two orbits leads to that the
SC gaps of the two orbits are out of synchronization when
approaching the critical points.
For example, in the situation with $R=0.3$ and $E_{\Delta}=0$, due
to the asymmetry of the two hopping integrals, the electrons in the
narrow orbit feel weaker attractive interaction than those in the
wide orbit, and forming a small SC gap. Thus, with the increasing of
$\delta$, the electrons in the narrow orbit will first enter the
normal state; at the same time, the electrons in the wide orbit may
be still SC, as seen in Fig.1b and 1c.

\begin{figure}[tp]
\vglue -0.6cm \scalebox{1.150}[1.15]
{\epsfig{figure=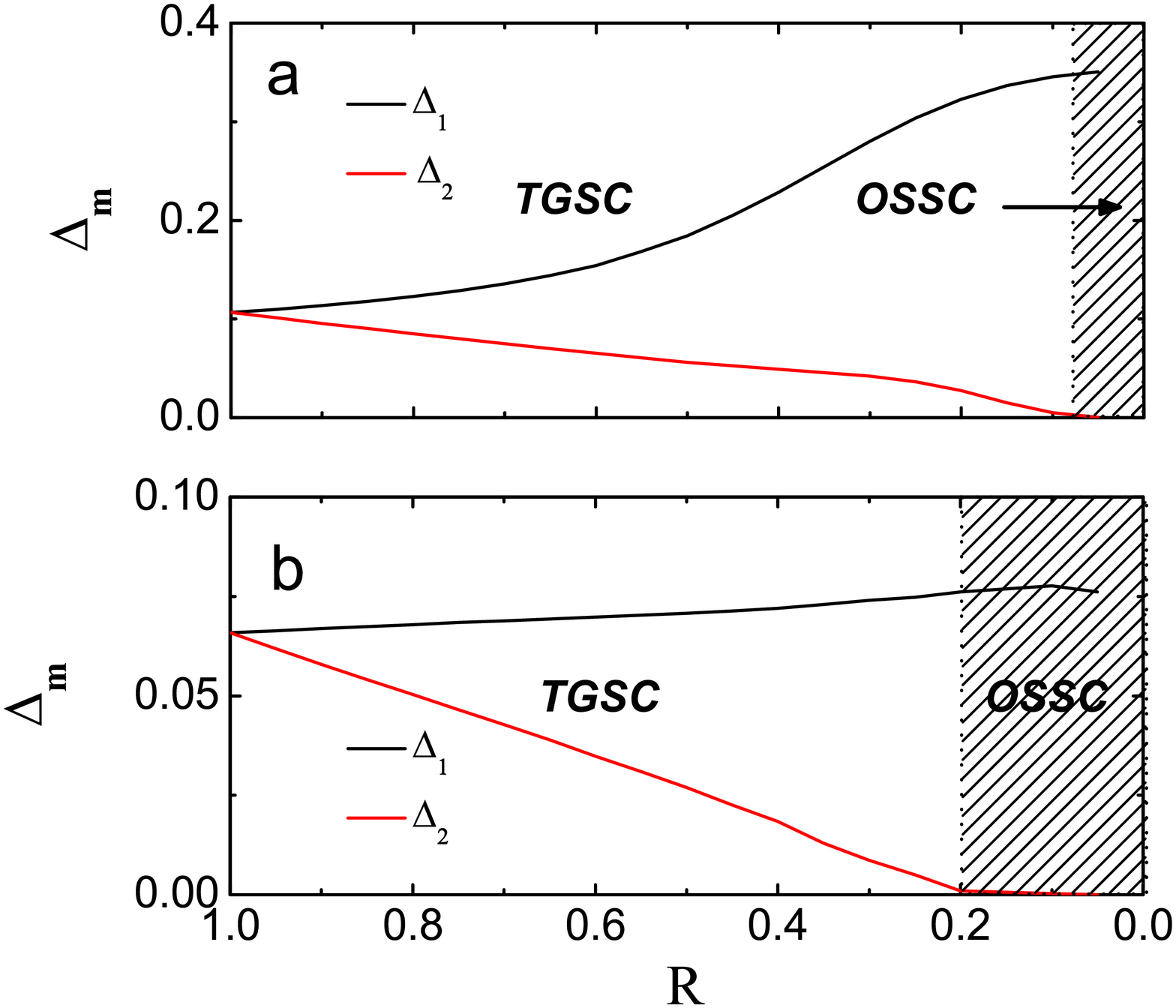,width=8cm,angle=0}} \caption{(Color
online) The order parameters $\Delta_{m}$ (m=1,2) vs ratio of
hopping integrals $R$ for different situations (a). $\delta=0.01$
and$\Delta=0$, and (b). $\delta=0.02$ and$\Delta=0$, respectively. }
\label{fig:fig2}
\end{figure}

To get more insight to the behavior of the OSSC phase, we study how
the phase diagram evolves with $R$, and the numerical result is
shown in Fig. 2. In the system with $\delta=0.01$ and
$E_{\Delta}=0$, it is clearly seen in Fig.2a that the difference
between $\Delta_{1}$ and $\Delta_{2}$ increases with the deviation
of $R$ from the unity. The SC order parameters exhibit different
behavior: $\Delta_{2}$ monotonously increases and saturates at
$R_{c}=0.08$; however, $\Delta_{1}$ monotonously decreases and
vanishes at $R_{c}$, indicating the appearance of the OSSC phase.
As the doping concentration increases to $\delta=0.02$ in Fig.2b,
the two SC order parameters behave similar to the first situation.
Finally, the TGSC-OSSC phase transition occurs at $R_{c}=0.22$.
The critical value of the TGSC-OSSC phase transition becomes larger
with the increase of $\delta$, in agreement with the results in Fig.
1.
The asymmetric behavior of the SC order parameters arises from the
symmetry breaking of the system in orbital space. Again, the
asymmetry of the electron kinetic energy in two orbits favors the
lift of the SC gap in the TGSC phase.
When the hopping integral ratio R is greater than the unity, the
behavior of $\Delta_{1}$ is inter-changed with that of $\Delta_{2}$.
%
%
%
%

\begin{figure}[tp]
\vglue -0.6cm \scalebox{1.150}[1.15]
{\epsfig{figure=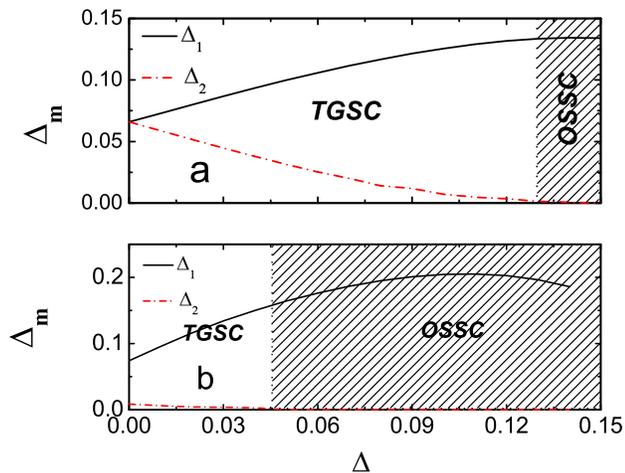,width=8cm,angle=0}} \caption{(Color
online) Dependence of the SC order parameters $\Delta_{m}$ (m=1,2)
on the level splitting $E_{\Delta}$ for (a). $R=1$ and
$\delta=0.02$, and (b). $R=0.3$ and $\delta=0.02$, respectively. }
\label{fig:fig3}
\end{figure}

\subsection{Crystal Field Splitting Dependence of SC Phases}
Not only the asymmetric hopping integrals, but also the crystalline
field splitting $E_{\Delta}$ can break the orbital SU(2) symmetry of
the system, leading to the formation of the OSSC phase, as seen in
Fig.3.
For the situation with $R=1$ and $\delta=0.02$ shown in Fig.3a, the
positive crystalline field splitting $E_{\Delta}$ enhances the SC
order parameter in orbit-1, $\Delta_{1}$, however, suppresses that
in orbit-2, $\Delta_{2}$. The system is in the TGSC phase when
$E_{\Delta}$ is lower than the critical value,
$E_{\Delta}^{c}$=0.125.
When the crystalline field splitting is greater than a critical
value, $E_{\Delta}^{c}$=0.125, the order parameter $\Delta_{2}$
vanishes. The system enters the OSSC regime.
For the situation with $R=0.3$, the phase diagram is shown in
Fig.3b. Compared with the situation in Fig.3a, the difference
between the SC order parameters $\Delta_{1}$ and $\Delta_{2}$ is so
significant that $\Delta_{2}$ is negligible as $E_{\Delta} >$ 0.042,
implying that the OSSC phase is more robust in the presence of the
large level splitting and the orbital-rotational symmetry breaking.

For the situation with R=1, the negative crystalline field splitting
E$_{\Delta}$ reverse the behavior of $\Delta_{1}$ and $\Delta_{2}$.
We find that similar to the role of the rotation symmetry breaking
in orbital space, the remove of the orbital level degeneracy by the
crystalline field splitting also enhances the occurrence of the OSSC
phase. Obviously, both of the cases break the symmetry of the
orbital space. Therefore, the asymmetry of the two orbits favors the
OSSC phase.

\begin{figure}[tp]
\vglue -0.6cm \scalebox{1.150}[1.15]
{\epsfig{figure=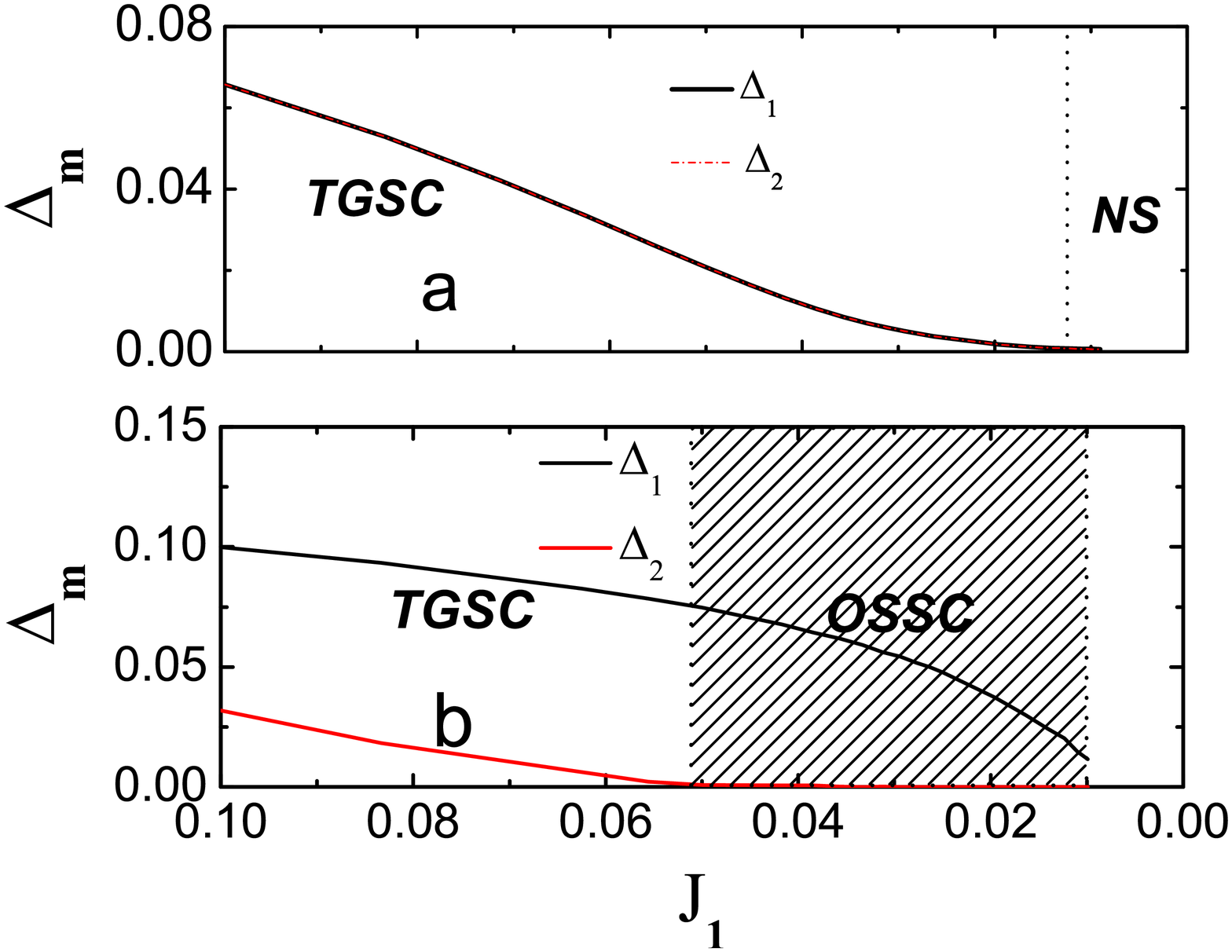,width=8cm,angle=0}} \caption{(Color
online) The SC order parameters $\Delta_{m}$ (m=1,2) vs the
superexchange coupling $J_{1}$ for different situations with $R=1$
and $\Delta=0$ (a), and $R=1$ and$\Delta=0.05$ (b), respectively, at
$\delta=0.02$. }
\label{fig:fig4}
\end{figure}

\subsection{Superexchange Coupling Dependence}
Similar to the single-orbital $t-J$ model \cite{Kotliar}, when two
superexchange pairing interactions $J_{1}$ and $J_{2}$ are too
small, the normal state is stable against the SC phase, as shown in
Fig.4a and Fig.4b.
In the orbital SU(4) system with $R=1$ and $E_{\Delta}=0$, the SC
order parameter of the orbit-1, $\Delta_{1}$, is equal to that of
the orbit-2, $\Delta_{2}$, due to the symmetry in the orbital space.
With the increase of the interaction strength $J_{1}$, the order
parameters $\Delta_{m}$ (m=1,2) become finite at $J_{1c}=0.013$. The
system lies in the TGSC regime.
Under the situation of $R=1$ and $E_{\Delta}=0.05$, the order
parameter $\Delta_{1}$ in the orbit-1 deviates from that in the
orbit-2 due to the breaking of the orbital SU(2) symmetry.
Compared with the case of the orbital SU(2) symmetry, a small
crystal field splitting drives the narrow-band SC order parameter
$\Delta_{2}$ to zero rapidly, and induces the occurrence of the
OSSC-normal phase transition near the critical value $J_{1c}=0.05$.
%
%

%
From the discussion above, we have demonstrated that in proper
parametric and doping regions, the OSSC phase is stablest, in
comparison with the TGSC phase and the normal state in the
multi-orbital $t-J$ model.
The requirement condition for the occurrence of the OSSC phase is
that the orbital SU(2) symmetry of the system is broken in the
orbital space. Generally speaking, the larger the hopping integral
ratio $R$ deviates from the unity, the larger the difference between
two gaps is; so does the crystalline field splitting $E_{\Delta}$.
Further, only in the proper doping and the interaction $J_{1}$
region, where the spin-fluctuation mediated pairing glue is strong
enough, the system is stable in the OSSC state.
One should recall that the present $s$-wave pairing symmetry is
$cos(k_{x})+cos(k_{y})$-type, which exhibits nodes along the
$k_{x}$=$k_{y} \pm \pi/2$, rather than the full-gapped BCS-type
$s$-wave SC. Such a singlet pairing symmetry is also in agreement
with Kubo's result \cite{Kubo}

The novelty of OSSC phase is the coexistence of the low-energy
"normal state" excitations and superconductivity. Such character
could be manifested directly in the tunneling experiments, $i.e.$,
the tunneling spectra may consist of those of the Cooper pairs and
the normal electrons.
To a certain extent, the OSSC phase is similar to the conventional
SC with nodes, such as the $d$-wave SC cuprate, or the $p$-wave SC
Sr$_2$RuO$_4$, $et$ $al.$
However, the OSSC phase differs from these states both
quantitatively and qualitatively. Quantitatively, the DOS of OSSC
phase is larger than those of the gapless modes; and qualitatively,
the OSSC phase does not have the pairing symmetry of the $d$-wave or
the $p$-wave SC.
Although the OSSC state bears the resemblance to the interior gap
superfluidity \cite{Liu}, there is qualitative difference between
these two states. Firstly, the microcosmic mechanisms of the
superconductivity are different, since the present SC phases are
mediated through the spin fluctuations, while the interior
superfluid forms through the Bose-Einstein condensation.
Secondly, the precondition of the interior gap superfluidity
requires that the effective masses of the quasiparticle in the two
branches is different; whilst, our theory predicts that the OSSC
phase can exist even if the effective masses in both orbits are
identical, providing that crystalline field splitting is large.
%
%

To date, no direct experimental observation about the novel OSSC
phase is available. Nevertheless, we could find some hints in the
anomalous properties of some unconventional SC.
Recently, by measuring the thermal conductivity and the specific
heat in the heavy-fermion SC Ce$_{1-x}$La$_{x}Co$In$_5$
\cite{Tanatar}, Tanatar $et$ $al.$ proposed that in the doped
compound, there coexist uncondensed electrons and nodal
quasiparticles. And more recent thermal measurement \cite{Flouquet}
demonstrated that in undoped CeCoIn$_5$, there exists the multigap
structure in the SC phase. From the present OSSC scenario, these two
behaviors are consistent with each other, rather than contradict
with each other. Since the number of orbits and the dispersion
relation in CeCoIn$_5$ are different from the present simple model,
more effort is needed to directly compare the present theory and
experimental results in CeCoIn$_5$.

Furthermore, the Fe-based SC discovered recently may be another
candidate of the OSSC phase. Some recent studies suggested that in
undoped LaOFeAs, the electron correlation between Fe 3d electrons is
strong and plays an important role in the ground state
\cite{Kotliar2, Laad}, and the first-principles electronic structure
calculations suggest that two or more orbits are involved in the
superconductivity, implying that the multi-orbital $t-J$ model is
appropriate for describing the low-energy physics of the iron-based
SC. With the increase of F-doping concentration, the system
undergoes from the normal to the SC states. The two-gap character in
sufficient F-doped LaOFeAs \cite{Mandrus, Zhenggq} suggests that in
the some doping region, these may exist the unconventional SC phase.
%
%
%
Surely, we expect that more elaborate experiments and the
comparisons between the theory and the experiment can be performed
to uncover the unconventional SC phase.

\section{Conclusions}

In summary, by using the extended auxiliary-boson approach, we have
demonstrated that in the multi-orbital $t-J$ models, besides the
two-gap superconducting phase, an orbital selective superconducting
ground state may be stable, when the orbital SU(2) symmetry is
broken in the correlated electronic systems. Such a new phase is
s-wave like.
The superconducting order parameters strongly depend on the
asymmetry of the hopping $R$ and the crystal field splitting
$E_{\Delta}$. The more the deviation from the orbital SU(2) symmetry
is, the more robust the orbital selective superconducting phase is.
Of course, the complicated dispersions relation of the multi-orbital
systems in realistic compounds may lead to more interesting
phenomena, and deserve further extensive investigation.

\acknowledgments

This work was supported by the NSF of China, the BaiRen Project and
the Knowledge Innovation Program of Chinese Academy of Sciences.
Part of the calculations were performed in Center for Computational
Science of CASHIPS and the Shanghai Supercomputer Center.


\begin{thebibliography}{}

\bibitem{Lee0}
P. A. Lee, N. Nagaosa, and X. G. Wen, Rev. Mod. Phys. {\bf 78}, 17 (2006).

\bibitem{Saxena}
S. S. Saxena, P. Agarwal, K. Ahilan, F. M. Grosche, R. K. W. Haselwimmer,
 M. J. Steiner, E. Pugh, I. R. Walker, S. R. Julian, P. Monthoux,
 G. G. Lonzarich, A. Huxley, I. Sheikin, D. Braithwaite, and J. Flouquet,
 Nature {\bf 406}, 587 (2000).

\bibitem{Huxley}
A. Huxley, I. Sheikin, E. Ressouche, N. Kernavanois, D. Braithwaite,
 R. Calemczuk, and J. Flouquet, Phys. Rev. B {\bf 63}, 144519 (2001).

\bibitem{Aoki}
D. Aoki, A. D. Huxley, E. Ressouche, D. Braithwaite, J. Flouquet,
 J. P. Brison, E. Lhotel, and C. Paulsen, Nature {\bf 413}, 613 (2001).

\bibitem{Yang}
H. Yang, X. Y Zhu, L. Fang, G. Mu, and H. H. Wen, cond-mat/08030623 (2008)


\bibitem{Maeno}
  Y. Maeno, H. Hashimoto, K. Yoshida, S. Nishizaki, T. Fujita,
  J. G. Bednorz, and F. Lichtenberg,
  Nature \textbf{372}, 532 (1994).

\bibitem{Mackenzie}
  A. P. Mackenzie and Y. Maeno,
  Rev. Mod. Phys. \textbf{75}, 657 (2003).

\bibitem{Yokoya} T. Yokoya et al, Science {\bf 294}, 2518 (2001).

\bibitem{Boaknin} E. Boaknin et al, Phys. Rev. Lett. {\bf 90}, 117003 (2003).

\bibitem{Canfield} P. C. Canfield and G. W. Crabtree,
Physics Today, {\bf 56}, No.3, 34 (2003).

\bibitem{Blumberg} G. Blumberg, A. Mialitsin, B. S. Dennis, N. D. Zhigadlo,
and J. Karpinski, Physica C {\bf 456} pp. 75-82 (2007).

\bibitem{Iavarone}
M. Iavarone, G. Karapetrov, A. E. Koshelev, W. K. Kwok, G. W. Crabtree,
and D. G. Hinks, cond-mat/0203329 (2002).

\bibitem{Tanatar}
M. A. Tanatar, Johnpierre Paglione, S. Nakatsuji, D. G. Hawthorn,
 E. Boaknin, R. W. Hill, F. Ronning, M. Sutherland, Louis Taillefer,
 C. Petrovic, P. C. Canfield, Z. Fisk,
Phys. Rev. Lett. {\bf 95}, 067002 (2005)

\bibitem{Flouquet}
G. Seyfarth, J. P. Brison, G. Knebel, D. Aoki, G. Lapertot and J.
Flouquet, Phys. Rev. Lett. {\bf 101}, 046401 (2008).

\bibitem{Boyer}
M. C. Boyer, W. D. Wise, Kamalesh Chatterjee, Ming Yi,
Takeshi Kondo, T. Takeuchi,
 H. Ikuta,  and  E. W. Hudson, Nature Physics {\bf 3}, 802 - 806 (2007)

\bibitem{Lee}
W. S. Lee, I. M. Vishik, K. Tanaka, D. H. Lu, T. Sasagawa, N. Nagaosa,
 T. P. Devereaux, Z. Hussain, and  Z.-X. Shen,
Nature {\bf 450}, 81 (2007).

\bibitem{Sawatzky}
S. Hufner, M. A. Hossain, A. Damascelli and G. A. Sawatzky, Rep.
Prog. Phys. {\bf 71}, 062501 (2008).

\bibitem{Hosono}
H. Takahashi, K. Igawa, K. Arii, Y. Kamihara, M. Hirano and H.
Hosono, Nature {\bf 453}, 376 (2008); X. H. Chen, T. Wu, G. Wu, R.
H. Liu, H. Chen and D. F. Fang, Nature {\bf 453}, 761 (2008).

\bibitem{Mandrus}
F. Hunte, J. Jaroszynski, A. Gurevich, D.C. Larbalestier, R. Jin,
A.S. Sefat, M.A. McGuire, B.C. Sales, D.K. Christen, D. Mandrus,
Nature, {\bf 453}, 903 (2008).

\bibitem{Zhenggq}
K. Matano, Z.A. Ren, X.L. Dong, L.L. Sun, Z.X. Zhao, Guo-qing Zheng,
arXiv:0806.0249v1.

\bibitem{Liu2}
A. Y. Liu, I. I. Mazin, and J. Kortus,
Phys. Rev. Lett. {\bf 87}, 87005 (2001).

\bibitem{Barzykin}
V. Barzykin and L. P. Gor'kov, cond-mat/0606191 (2006)

\bibitem{Agterberg}
D. F. Agterberg, T. M. Rice, and M. Sigrist,
Phys. Rev. Lett. {\bf 78}, 3374 (1997).

\bibitem{Anisimov} V. Anisimov, I. Nekrasov, D. Kondakov, T. Rice and M.
Sigrist, Eur. Phys. J. B \textbf{25}, 191, (2002).

\bibitem{Nakatsuji} S. Nakatsuji and Y. Maeno,
Phys. Rev. Lett. \textbf{84}, 2666 (2000).

\bibitem{Lee2} J. S. Lee, S. J. Moon, T.W. Noh, S. Nakatsuji, and Y. Maeno,
Phys. Rev. Lett. \textbf{96}, 057401 (2006).

\bibitem{Wang} S.-C. Wang et al., Phys. Rev. Lett. \textbf{93}, 177007
(2004).

\bibitem{Dai} X. Dai et al., cond-mat/0611075v1
(2004).

\bibitem{Imada}
  M. Imada, A. Fujimori, and Y. Tokura,
  Rev. Mod. Phys. {\bf 70}, 1039 (1998).

\bibitem{Santini}
  P. Santini, R. L\'{e}manski, and P. Erd\"{o}s,
  Adv. Phys. \textbf{48}, 537 (1999).

\bibitem{Liu}
  W. Vincent Liu, and Frank Wilczek, Phys. Rev. Lett. {\bf 90}, 047002 (2003).

\bibitem{Takimoto}
  T. Takimoto, Phys. Rev. B \textbf{62}, R14641 (2000).

\bibitem{Takimoto2}
  T. Takimoto, T. Hotta, and K. Ueda,  Phys. Rev. B \textbf{69}, 104504 (2004).

\bibitem{Mochizuki}
  M. Mochizuki, Y. Yanase, and M. Ogata,
  Phys. Rev. Lett. \textbf{94}, 147005 (2005).

\bibitem{Kubo}
  K. Kubo, Phys. Rev. B {\bf 75}, 224509 (2007).

\bibitem{Anderson}
P. W. Anderson, Nature {\bf 235}, 1196 (1987).

\bibitem{Baskaran}
G. Baskaran, Z. Zou, P. W. Anderson, Solid State Commun, {\bf 63},
973 (1987).

\bibitem{Castellani}
C. Castellani, C. R. Natoli and J. Ranninger, Phys. Rev. B. {\bf 18},
4945 (1978).

\bibitem{Lu1}
Feng Lu, Wei-Hua Wang and Liang-Jian Zou, Phys. Rev. B. {\bf 77},
125117 (2008).

\bibitem{Lu2}
Feng Lu, Dong-meng Chen and Liang-Jian Zou, cond-mat/0605379

\bibitem{Fujii}
T. Fujii, Y. Tsukamoto, and N. Kawakami, cond-mat/9811189

\bibitem{Schlottmann}
P. Schlottmann, Phys. Rev. Lett. {\bf 69}, 2396 (1992).

\bibitem{Kotliar}
G. Kotliar, and J, Liu, Phys. Rev. B. {\bf 38}, 5142 (1988).

\bibitem{Coleman}
P. Coleman, Phys. Rev. B. {\bf 29},
3035 (1984), ibid  {\bf 35}, 5072 (1987)

\bibitem{Valatin}
D. G. Valatin, Nuovo Cimento {\bf 7}, 843 (1958).

\bibitem{Callaway}
Joseph Callaway, Quantum theory of the solid state,
Academic Press, (New York , 1976)

\bibitem{Kotliar2} K. Haule, J. H. Shim and G. Kotliar,
Phys. Rev. Lett. 100, 226402 (2008)

\bibitem{Laad} L. Craco, M. S. Laad, S. Leoni and H. Rosner,
arXiv:0805.3636v1


\end{thebibliography}
\end{document}